\documentclass{iau} 
\usepackage{graphicx}

\usepackage[utf8]{inputenc}
\usepackage{bm}
\usepackage{amsmath}
\usepackage{amssymb}
\usepackage{tensor}
\usepackage{natbib}
\usepackage{siunitx}
\usepackage{color}

\newcommand{\arcmancer}{\textsc{Arcmancer}}
\newcommand{\Msun}{\text{M}_{\odot}}
\newcommand{\vct}[1]{\bm{\mathit{#1}}}
\newcommand{\mat}[1]{\bm{\mathrm{#1}}}
\newcommand{\derfrac}[2]{\frac{\ud #1}{\ud #2}}
\newcommand{\mulmat}{\mat{M}} 
\newcommand{\svec}{\vct{I}} 
\newcommand{\evec}{\vct{J}} 
\newcommand{\cmulmat}{\bm{\mathcal{M}}} 
\newcommand{\csvec}{\bm{\mathcal{I}}} 
\newcommand{\cevec}{\bm{\mathcal{J}}} 
\newcommand{\ud}{\mathrm{d}} 
\newcommand{\afpar}{\lambda}
\newcommand{\rsfac}{\mathcal{G}}
\newcommand{\chr}{\Gamma}

\newcommand{\ix}[1]{\indices{#1}}
\newcommand{\TM}{TM}
\newcommand{\coTM}{T^{*}M}
\newcommand{\fR}{\mathbb{R}}
\newcommand{\fromto}{\rightarrow} 

\makeatletter
\let\jnl@style=\it
\def\ref@jnl#1{{\jnl@style#1}}

\def\aap{\ref@jnl{A\&A}}                
\def\apss{\ref@jnl{Ap\&SS}}             
\def\mnras{\ref@jnl{MNRAS}}             
\def\apj{\ref@jnl{ApJ}}                 
\def\apjl{\ref@jnl{ApJ} (Letters)}                

\makeatother

\title[Relativistic polarized radiative transfer] 
{Ray-tracing and polarized radiative transfer in General Relativity}

\author[Pauli Pihajoki, Antti Rantala \& Peter H. Johansson]   
{Pauli Pihajoki$^1$, Antti Rantala$^1$ \& Peter H.~Johansson$^1$
}

\affiliation{$^1$University of Helsinki, Department of Physics, \\
    Gustaf H\"allstr\"omin katu 2a, 00560 Helsinki, Finland \\
    email: {\tt pauli.pihajoki@iki.fi}
}

\pubyear{2016}
\volume{324}  
\jname{New Frontiers in Black Hole Astrophysics}
\editors{A.C. Editor, B.D. Editor \& C.E. Editor, eds.}
\begin{document}

\maketitle

\begin{abstract}
We discuss the problem of polarized radiative transfer in
general relativity. We present a set of equations suitable for solving
the problem numerically for the case of an arbitrary space-time metric,
and show numerical solutions to example problems. The solutions
are computed with a new ray-tracing code, \arcmancer{}, developed by the
authors.

\end{abstract}

\firstsection 
\section{Introduction}

In curved space-times, light is observed to propagate along curved paths.
As such, computing mock observations of scenes affected by strong
gravity requires taking general relativity fully into account. A conceptually
simple approach in the limit of geometric optics is to compute
the paths of light, given by null geodesics. This can be done numerically or
analytically, in the case of highly symmetric space-times such as the
Kerr space-time.
This approach has been
often used in the literature since the late 1960's
\citep[e.g.][]{cunningham1973,luminet1979,dexter2016}.
A complete
solution of the polarized radiative transfer problem in curved space
requires not only solving the geodesic, but solving the general relativistic
polarized radiative transfer equation along this geodesic as well.

Currently, there are several codes capable of computing general
relativistic polarized radiative transfer: \textsc{grtrans}
\citep{dexter2016}, \textsc{Astroray} \citep{shcherbakov2013} and the
code described by \citet{broderick2003b,broderick2003a}.
However, none of
codes above support arbitrary space-time metrics, and are instead
restricted to the Kerr metric.
To remedy this, we have developed \arcmancer{} (Pihajoki et
al. 2017, \emph{in prep}): a
C++ library for computing geodesics and polarized radiative transfer in
space-times with arbitrary user defined metrics. In the following, we
will briefly describe the general relativistic polarized radiation
transfer problem and show some promising initial results.

\section{Polarized radiative transfer in curved space-time}

To compute radiative transfer along the path of light, given by the
null geodesic $\gamma$, we need to first solve it numerically.
A geodesic can be given as a curve
$\gamma:\fR\supseteq I \fromto M$, where $M$ is the space-time manifold,
which satisfies the equations of motion. The equations of motion can
be given in two complementary forms. For initial conditions
$(x,k^a)\in\TM$, where $x\in M$ is the initial position and $k^a\in\TM_{x}$
is the initial tangent vector of the geodesic, given on the tangent
bundle, we have in a coordinate frame
\begin{align}\label{eq:eqmolag}
    \dot{x}^\alpha &= k^\alpha \\
    \dot{k}\ix{^\alpha} &=
    -g\ix{^{\alpha\beta}}\chr\ix{_{\alpha\beta\delta}}k\ix{^\beta}k\ix{^\delta} \\
    \chr\ix{_{\alpha\beta\delta}} &= \frac{1}{2}\left(
        \partial_{\beta}g\ix{_{\alpha\delta}}
        + \partial_{\delta}g\ix{_{\alpha\beta}}
        - \partial_{\alpha}g\ix{_{\beta\delta}}
    \right)
\end{align}
where we define $\dot{x} = \ud x/\ud\afpar$,  and where $\afpar$ is the
affine parameter of the geodesic, $g\ix{_{ab}}$ is the metric and
$\chr\ix{_{\alpha\beta\delta}}$
are the Christoffel symbols of the first kind. For initial conditions
$(x,p_a)\in\coTM$, where $p_a\in\coTM_{x}$ is the initial four-momentum of
the geodesic, given on the cotangent bundle, we can define a
Hamiltonian $H$ and the Hamiltonian equations of motion a coordinate
frame as
\begin{align}
    H(x,p_a) &= \frac{1}{2}g(x)\ix{^{bc}}p\ix{_b}p\ix{_c} 
    \quad\text{($= 0$, for null geodesics)}\label{eq:hamiltonian} \\
    \dot{x}\ix{^\alpha} &= \frac{\partial H}{\partial p\ix{_\alpha}} = g(x)\ix{^{\alpha\beta}}p\ix{_\beta}\\
    \dot{p}\ix{_\alpha} &= -\frac{\partial H}{\partial x\ix{^\alpha}} 
            = -\frac{1}{2}\partial_{\alpha}g(x)\ix{^{\beta\delta}}p\ix{_\beta}p\ix{_\delta}.
\end{align}
The two different parametrizations are connected by the natural bijection
$p\ix{_a} = g\ix{_{ab}}k\ix{^b}$. However, it should be noted that when plasma
effects cannot be ignored, the path of light is not given by a geodesic
following the dispersion relation \eqref{eq:hamiltonian}, but a
dispersion relation depending on the polarization and local properties
of the plasma \citep{broderick2003a}. These effects are not important
for the applications presented in this paper, so we can safely ignore
them here.

To compute polarized radiative transfer along a geodesic, we first need
to fix a polarization frame along each point of the geodesic, and
connect these frames to the polarization frame at the point of
observation. Conceptually the simplest method is to define a frame
of two four-vectors $\{h^a,l^a\}$, representing local horizontal and
vertical directions, in the spatial subspace of the
observer, who is assumed to move with a four-velocity $u^a$. If $h^a$ and
$l^a$ are orthogonal with respect to each other as well as with respect to
to the tangent vector of the geodesic, $k^a$, they define a proper polarization frame. 
This frame can then be
parallel propagated along the geodesic to obtain a well defined
polarization frame at each point of the geodesic.
Next, the redshift $\rsfac$ between a
point on the geodesic and the observation point is given by
\begin{equation}\label{eq:gfactor}
    \rsfac = \frac{\nu_0}{\nu} 
    = \frac{{u_0}\ix{^a}k(x_0)\ix{_a}}{{u}\ix{^a}k(x)\ix{_a}},
\end{equation}
where $k^a$ is the tangent vector of the geodesic, $\nu_0$ and $\nu$ are
the observed and emitted frequencies, $u_0^a$ and $u^a$ are the observer
and emitter four-velocities,  and $x_0$ and $x$ are the observation and
emission points, respectively. Finally, we need a relativistic
generalization of the flat space radiative transfer equation
\begin{equation}\label{eq:newtonian_rt}
\derfrac{\svec_\nu}{s} = \evec_\nu - \mulmat\,\svec_\nu,
\end{equation}
where $s$ is the physical distance through the medium,
$\svec = (I,Q,U,V)$ is the Stokes vector,
$\evec = (j_I, j_Q, j_U, j_V)$ is the vector of corresponding emissivities and
\begin{equation}
    \mulmat =
    \Biggl(
    \begin{smallmatrix}
        \alpha_I & \alpha_Q & \alpha_U & \alpha_V \\
        \alpha_Q & \alpha_I & r_V      & -r_U \\
        \alpha_U & -r_V     & \alpha_I & r_Q \\
        \alpha_V & r_U      & -r_Q     & \alpha_I
    \end{smallmatrix}
    \Biggr)
\end{equation}
is the M\"uller matrix containing the absorption coefficients $\alpha_i$ and
the Faraday conversion and mixing coefficients $r_i$ for the various
Stokes parameters.
When the four-velocity of the local medium
(i.e.\ astrophysical  plasma) and a polarization reference direction $b^a$
(given e.g.\ by the local direction of the magnetic field $B$) 
is specified, we can use the parallel
propagated polarization frame of the observer to
generalize equation~\eqref{eq:newtonian_rt}, yielding
\begin{align}
    \derfrac{\csvec_{\nu_0}}{\afpar} &= \cevec_{\nu_0} - \cmulmat_{\nu_0}\csvec_{\nu_0}
    \text{, where} \label{eq:grpoltrans} \\
    \csvec_{\nu_0}         &= \frac{\rsfac^3}{\nu_0^3} \svec_{\nu_0/\rsfac},
    \quad\quad
    \cevec_{\nu_0}         = \frac{\rsfac^2}{\nu_0^2} R(\chi)\evec_{\nu_0/\rsfac}, \\
    \cmulmat_{\nu_0} &= \frac{\rsfac}{\nu_0} R(\chi)\mulmat_{\nu_0/\rsfac}R(-\chi) 
    \quad\text{and}\quad
    R(\chi) =
    \Biggl(
    \begin{smallmatrix}
        0 & 0 & 0 & 0 \\
        0 & \cos(2\chi) & -\sin(2\chi)  & 0 \\
        0 & \sin(2\chi) & \cos(2\chi)  & 0 \\
        0 & 0 & 0 & 0
    \end{smallmatrix}
    \Biggr),
\end{align}
which are in general functions of frequency $\nu$, the angle
$\theta$ between the four-velocity $k^a$ of the geodesic and the local
reference direction $b^a$, and the angle $\chi$ between the parallel
propagated polarization frame and the local reference polarization
frame, projected orthogonal to $k^a$. See \citet{shcherbakov2011} for details.
Given initial conditions for the intensity $\csvec$, the emissivity $\evec$,
the response tensor $\mulmat$, the local four-velocity $u^a$ and the local
polarization reference direction $b^a$ on all points of a geodesic,
equation \eqref{eq:grpoltrans} can be numerically integrated along the
geodesic, yielding the solution to the general relativistic polarized
radiative transfer problem.

\section{Example problems and future plans}

\begin{figure}[htbp]
    \begin{center}
        \includegraphics[width=0.62\textwidth]{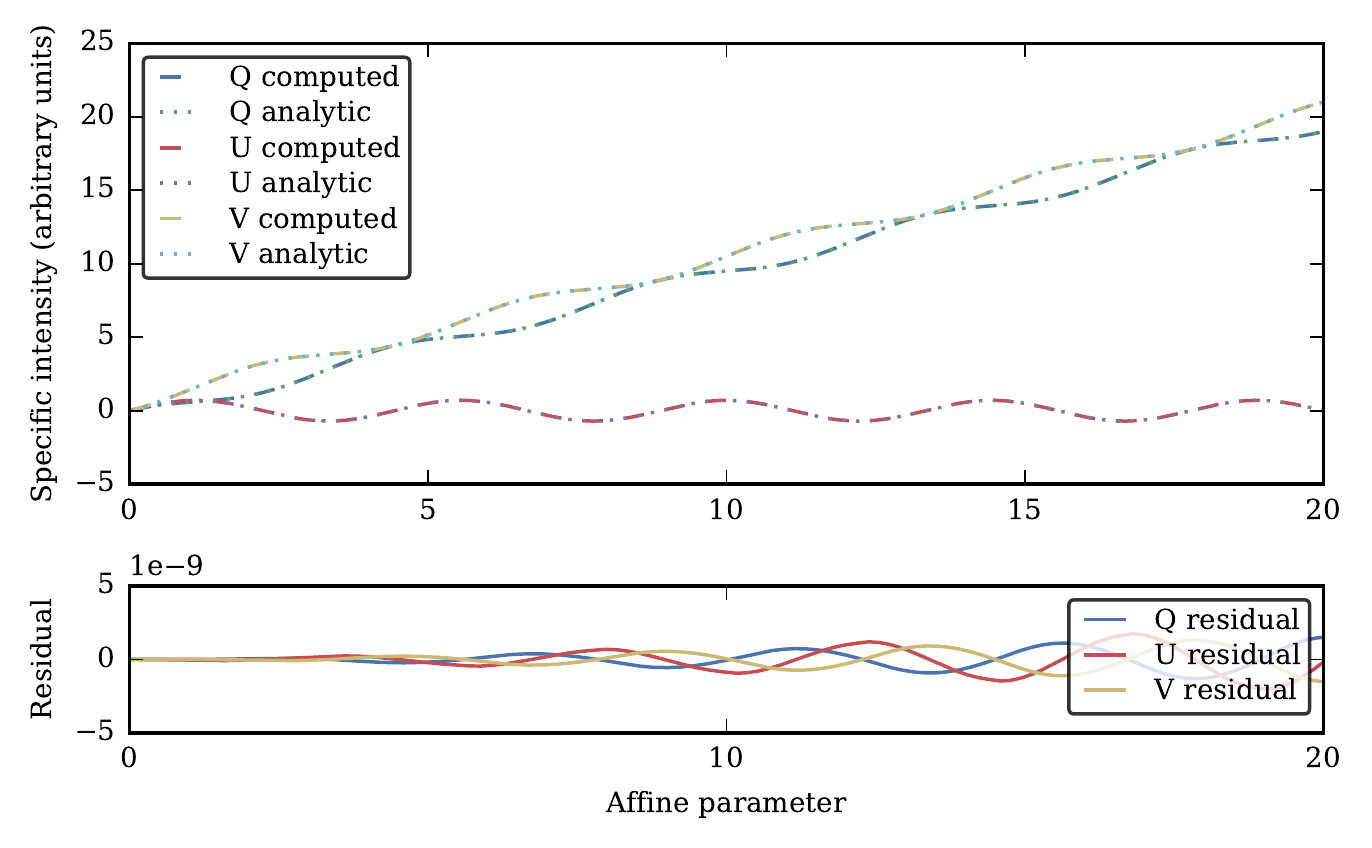}
        \vspace{-0.5cm}
        \caption{
            \label{fig:polcomp}
            Comparison of a numerical (points) and an analytic (dashed lines)
            solution to polarized radiative transfer with constant
            coefficients in Minkowskian metric. Upper panel show the
            solutions in arbitrary units and lower panel show the
            residuals between numerical and analytic solutions.
            The panels show the solution to a problem with
            constant emission $\cevec = (0,1,1,1)$ and Faraday
            rotation and conversion factors $r_V=r_Q=1$,
            with rest of the $\cmulmat$
            entries identically zero, starting with initial values
            $\csvec=(0,0,0,0)$.
            $I$ is zero to within numerical precision and is not shown.
            For the analytic solutions, see \citet{dexter2016}.
        }
        \end{center}
\end{figure}

\begin{figure}[htbp]
    \begin{center}
        \includegraphics[width=0.94\textwidth]{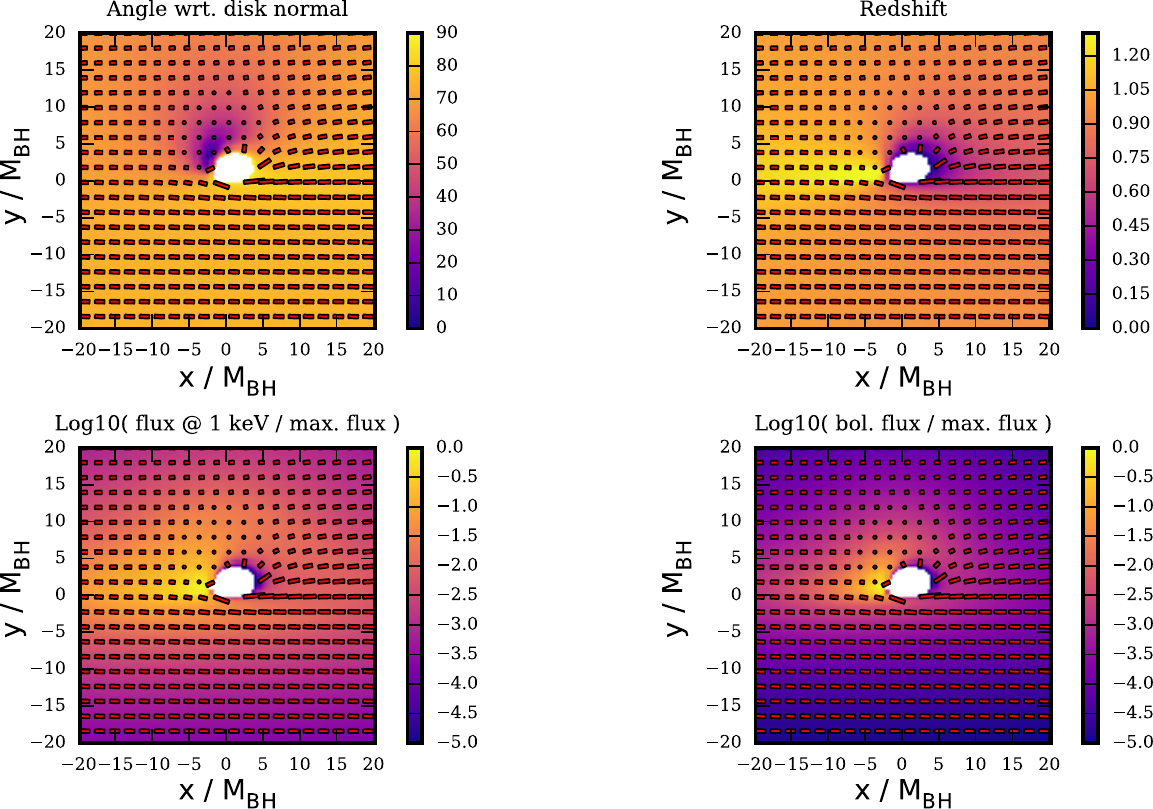}
        \caption{
            \label{fig:ntdisk}
            Ray-traced images of a Kerr black hole with a mass
            of $M=10\Msun$, accretion rate of $\dot{M}=0.1
            M_\text{Edd}$ and a dimensionless
            spin of $a/M = 0.998$. The surrounding accretion disc is
            optically thick and geometrically thin, modelled with the
            Novikov--Thorne equations \citep{ntdisk}. The emission is
            modelled with a semi-infinite electron scattering atmosphere
            \citep{chandrasekhar1960}. The direction of linear
            polarization is shown by the red bars, and the degree of
            polarization is indicated by the bar lengths, with
            the degree of polarization at the bottom edges corresponding
            to $\sim 4\%$.
            \emph{Top left}: The angle between the geodesic and the disc
            normal as measured by an observer co-moving with the disc
            matter.
            \emph{Top right}: The redshift factor, $\rsfac$, between the disc
            matter and the observation point
            \emph{Bottom left:} Observed flux at $1\,\si{keV}$,
            integrated over a $10\,\si{keV}$ window, relative to the
            maximum flux, in logarithmic units.
            \emph{Bottom right:} Observed bolometric flux, relative to
            the maximum flux, in logarithmic units.
        }
        \end{center}
\end{figure}

The equations described above have been implemented in the \arcmancer{}
library. Using the library we can compute results to some example
problems. Figure~\ref{fig:polcomp} shows the comparison between
numerical and analytic results to two examples of polarized radiative
transfer problems in Minkowskian flat space. We see that the analytic
results are well matched within the given numerical tolerance.
Figure~\ref{fig:ntdisk} shows numerically computed images of a
Novikov--Thorne accretion disc \citep{ntdisk} around a rotating Kerr
black hole. The relativistic effects on the direction of the linear
polarization and on the degree of polarization are clearly visible.

Our future plans are focused on testing the \arcmancer{} code with a larger and
more demanding set of applications, such as optically thin black hole accretion
flows and black hole jet synchrotron plasmas. Eventually, the full code
together with the Python interface will be made freely available for the general
public.

\end{document}